\documentclass [floatfix, superscriptaddress, twocolumn, showpacs, aps, pra, 10pt]{revtex4-1}
\usepackage{amsmath}
\usepackage{natbib}
\usepackage{hyperref}
\usepackage{graphicx}
\usepackage{bbm}
\usepackage{amssymb}
\usepackage[english]{babel}
\usepackage{lmodern}
\usepackage[T1]{fontenc}
\usepackage{wasysym}
\usepackage[normalem]{ulem}

\usepackage{dcolumn}
\usepackage{color}

\graphicspath{{Figs/}}

\DeclareMathOperator{\e}{\displaystyle e}

\usepackage{amssymb}

\begin{document}

\title{Harmonic fine tuning and triaxial spatial anisotropy of dressed atomic spins}


\author{Giuseppe Bevilacqua}
\email{giuseppe.bevilacqua@unisi.it}
\author{Valerio Biancalana}
\affiliation{Dept. of Information Engineering and Mathematics - DIISM, University of Siena -- Via Roma 56, 53100 Siena, Italy}
 \author{Antonio Vigilante}
\affiliation{Dept. of Information Engineering and Mathematics - DIISM, University of Siena -- Via Roma 56, 53100 Siena, Italy}
\affiliation{Department of Physics and Astronomy, University College London, Gower Street, London WC1E 6BT, United Kingdom}    
\author{Thomas Zanon-Willette}
\affiliation{Sorbonne Universit\'e, Observatoire de Paris, Universit\'e PSL, CNRS, LERMA, F-75005, Paris, France}
\author{Ennio Arimondo}
\affiliation{Dipartimento di Fisica E. Fermi, Universit\`a of Pisa -- Lgo. B. Pontecorvo 3, 56127 Pisa, Italy}
\affiliation{INO-CNR, Via G. Moruzzi 1, 56124 Pisa, Italy}

\date{\today}


 \begin{abstract}
   The addition of a weak oscillating field modifying strongly dressed
   spins enhances  and enriches  the system quantum  dynamics. Through
   low-order  harmonic   mixing  the  bichromatic   driving  generates
   additional rectified  static field acting  on the spin  system. The
   secondary field allows for a fine tuning of the atomic response and
   produces effects not accessible  with a single dressing field, such
   as a spatial triaxial anisotropy of the spin coupling constants and
   acceleration   of   the   spin   dynamics.    This   tuning-dressed
   configuration  introduces  an  extra  handle for  the  system  full
   engineering  in quantum  control applications.   Tuning amplitude,
   harmonic  content,  spatial  orientation  and  phase  relation  are
   control parameters.  A  theoretical analysis, based on perturbative
   approach,  is experimentally  tested by  applying  a bichromatic
   radiofrequency field  to an optically pumped Cs  atomic vapour.  The
theoretical  predictions  are   precisely  confirmed  by  measurements
performed with tuning frequencies up to the third harmonic.  

\end{abstract}

\date{\today}

\maketitle

Dressing of  a quantum system by a  non-resonant electromagnetic field
represents an important tool  within  quantum control. Energies and
electromagnetic response  are modified  by the dressing.   Seminal
work          of          Cohen-Tannoudji         and          Haroche
(CTH)~\cite{CohenTannoudjiHaroche_66,HarocheCohen_70_1}   derived  the
modifications of  the spin precession  frequency in a  static magnetic
field  in presence  of a  strong radiofrequency  (rf)  dressing field,
off-resonant and linearly polarised orthogonally to the static one.  A
key  dressing  signature  is  the  $J_0$  zero-order  Bessel  function
dependence of the eigenenergies.   The dressing produces as additional
feature  a cylindrical  spatial anisotropy  for the  evolution  of the
quantum   coherences~\cite{LandreCTH_70}.    The   $J_0$   eigenenergy
collapse    was   examined   for    atoms   in~\cite{Yabuzaki_jpsj_72,
  Kunitomo_pra_72,ItoYabuzaki_1994,Muskat1987,EslerTorgerson_07,Chu_11},
for a  Bose-Einstein condensate in  ~\cite{BeaufilsGorceix_08}, for an
artificial              atom              in~\cite{TuorilaHakonen_10}.
Ref.~\cite{bevilacqua_pra_12}  investigated  the  generalization to  a
dressing with a periodic  arbitrary waveform.  The close connection of
the  $J_0$ collapse  with the  tunneling suppression  was  pointed out
in~\cite{GrifoniHaenngi_98,Holthaus_pra_2001}, and  with the dynamical
localization     freezing     in     optical     lattices     reviewed
in~\cite{Eckardt_17}.   The dynamical driving  and  the $J_0$
Bessel   response   were   described   as   a   frequency   modulation
in~\cite{AshhabNori_07}, and  extended to the  presence of dissipation
in~\cite{HausingerGrifoni10}.    Critical   dressing   based  on   the
simultaneous dressing of two spin species to the same effective Larmor
precession              frequency             was             explored
in~\cite{HarocheCohenTannoudji_prl_70_2,GolubLamoreaux_94,Swank_pra_18}.
A variety of microwave and  rf dressings was explored in recent years,
with  those   based  on   the  $J_0$  response   for  cold   atoms  in
~\cite{gerbier_pra_06,  Hofferberth_pra_07},   for  a  two-dimensional
electron gas in~\cite{ previshko_prb_15}, for high resolution
magnetometry in~\cite{Swank_pra_18},  and for the control  of 
spin-exchange  relaxation in~\cite{Hao_pra_19}.  The  dressing applied
in~\cite{Haroche_71}  to  compensate  an  inhomogeneous 
distribution, was extended determining magic 
{dressing  parameters} based on corrections to  the $J_0$
response~\cite{ZanonArimondo_12}, or
applying an inhomogeneous dressing field~\cite{Bevilacqua_apl_19}. \\
\indent  This work  introduces a  flexible quantum  handle  allowing a
continuous  control between  collapse and  enhancement of  the quantum
response. The tuning  tool is a weak non-resonant  additional rf field
operating  in  the   split  biharmonic  driving  configuration,  i.e.,
oscillating  at a  low order  harmonic of  the dressing  frequency and
applied  along  a direction  orthogonal  to  the  dressing one.   This
configuration demonstrates performances unmatched by the 
single-harmonic  system. A  quantum coupling  more versatile  than the
$J_0$ dependence  and  a  triaxial  spatially
anisotropic response are the tuning-dressed signatures.
The tuning  interaction produces  a modification of  the eigenenergies
depending  on the  spatial direction  of the  applied  magnetic field,
namely an undressed response along  the dressing field direction and a
fully  tunable one  in the  orthogonal plane.  \\ 
\indent The introduction of a secondary field into a dressed system
produces a large and easily realized quantum enrichment 
in the preparation and manipulation  of the spin dynamics, leading
also  to a magnified  quantum response.   In the  quantum information
language,  our   quantum  handle  represents   an  additional  storage
resource.   The  tuning-dressed features  are  useful  to all  quantum
research  areas,  from simulation  to atomic  interferometry,
spintronics,   superconducting  circuits,  vacancy
centers, atomic clocks, in addition to magnetometry as
in this work.  A temporal modulation of the tuning
field  amplitude may  enlarge  the dynamical  driving  access for  the
qubits.   The  anisotropic  response   introduces  for  the  qubits  a
configuration  existing  in  systems  as the  ferromagnets.   For  the
anisotropy applications  in interferometry with  artificial or natural
atoms~\cite{OnoNori_19,AmitFolman_19}, a quantum tuning 
with a controlled collapse along different spatial directions
may  realize  large  area  Stern-Gerlach  spin-splitters  and  mirrors
leading to  a higher sensitivity.   Our approach has the  potential to
spatially modulate the spin  exchange interactions in ultracold spinor
mixtures, opening  up to quantum simulations  with tunable anisotropic
Heisenberg  interactions. That  anisotropy would  also offer  a new  rf compensation  of  the ac
tensorial   shift   modifying   the   optical   clock   operation   in
alkaline-earth  with nuclear spin  not equal  to one-half.   A tunable
triaxial spin response will offer a new control of spin currents in 2D
or 3D  condensed matter systems, modifying  spin-orbit interaction and
opening a dressed spintronics direction. 

  Our theory  is  based  on a  perturbative  treatment for  the
quantum coupling  to static  and tuning fields  of a  strongly dressed
quantum system,  not treated  within the rotating  wave approximation.
In  \cite{BevilacquaSuppl20} a  description appropriate  for  the high
spin  atomic system  explored in  the  experiment is  given.  Here  we
consider  a   spin  1/2  system  (either  real   or  artificial  atom)
interacting with  a static  magnetic field having  components $B_{0j}$
along the  $j=(x,y,z)$ axes.   For an atomic  system with  $g$ Land\'e
factor  and $\mu_B$  the  Bohr magneton,  the  spin-field coupling  is
determined by the gyromagnetic ratio $\gamma=g\mu_B$ and characterized
by the energies $\omega_{0j}=\gamma B_{0j}$, ($\hbar=1$).  The
system  is driven by  two magnetic  fields oscillating  at frequencies
$\omega$  and $\omega_t=p\omega$,  dressing  and tuning  respectively,
$B_d$ oriented along  the $x$ axis and $B_t$ along  the $y$ axis.  The
corresponding   Rabi  frequencies   are   $\Omega_d=\gamma  B_d$   and
$\Omega_t=\gamma B_t$.

Introducing the $\tau=\omega t$ time, the Hamiltonian is
\begin{equation}
H= \sum_{j=(x,y,z)}\frac{\omega_{0j}}{2 \omega}\sigma_j+\frac{\Omega_d}{2\omega}\cos(\tau)\sigma_x
+\frac{\Omega_t}{2\omega}\cos(p\tau+\Phi)\sigma_y,
\label{Hamilt:eq}
\end{equation}
where  $\sigma_j$  are  the   Pauli  matrices  and  $\Phi$  the  phase
difference between the two oscillating fields. 

Defining the $\xi=\Omega_d/\omega$  dressing parameter, we explore the
strong dressing with  $\xi\gg \Omega_t/\omega, \omega_{0j}/\omega$ for
all  $j$.  Within  the  perturbative analysis  we  factorize the  time
evolution  operator as  $U=U_0U_I$.  The $U_0$  dressing evolution  is
given by
\begin{equation} 
U_0=e^{-i\varphi(\tau)\sigma_x/2},
\end{equation}
where $\varphi(\tau)= (\Omega_d/\omega) \sin\left(\tau\right) $.  As
detailed in
\cite{BevilacquaSuppl20}
the $U_I$
interaction evolution  is given by the following equation:
\begin{equation}
  \label{eq:UI}
    i\dot{U}_I   = 
    \left[ \frac{\omega_{0x}}{2\omega}\sigma_x+g_y(\tau) \sigma_y+g_z(\tau)\sigma_z \right] U_I = \epsilon A(\tau) U_I,
\end{equation}
%
and $\epsilon$ is a
bookkeeper for the perturbation orders.

As  the $A$ matrix  is periodic, we use the Floquet theorem
to write
\begin{equation}
  \label{eq:flo:magnus}
  U_I(\tau) = \e^{-iP(\tau)} \, \e^{-i \Lambda \, \tau}
\end{equation}
with  $P(0)=0$  and  $P(\tau+2\pi)  = P(\tau)$.   The  Floquet  matrix
$\Lambda$   is  a  time-independent   matrix.     Applying     to    $U_I$    the    Floquet-Magnus
expansion~\cite{magnus,BukovPolkovnikov_15} and writing $P = \epsilon
P_1 + \ldots $ and $ \Lambda = \epsilon \Lambda_1 + \ldots$, 
%
%
%
we obtain
\begin{equation}
  \label{eq:F1}
  \Lambda_1 = \frac{1}{2 \omega }\mathbf{h}\cdot\boldsymbol{\sigma}.
\end{equation}
We introduce here the  effective rectified magnetic field $\mathbf{h}$
driving  the spin evolution.  For $p$  even, $\mathbf{h}$  measured in
energy units is 
\begin{equation}
  \label{eq:effectivefield}
  \mathbf{h}=
  \begin{pmatrix}
    \omega_{0x} \\
     J_0(\xi)\omega_{0y}+ J_p(\xi)\Omega_t\cos(\Phi)\\
    J_0(\xi)\omega_{0z}
  \end{pmatrix}. 
\end{equation}
For $p$ odd, the $J_p$ term is added to the $z$ component with
$\cos(\Phi)$  replaced by $\sin(\Phi)$.   The excitation  with several
harmonic frequencies  and arbitrary orientations for  the tuning field
presented in~\cite{BevilacquaSuppl20} leads to
an extended quantum 
control.  However it  does not  modify the  geometry of  the rectified
fields generated  in the $yz$  plane orthogonal to the  dressing field
direction.  We verify  that  the second  order perturbative  expansion
generates an extra effective field oriented along the direction of the
dressed field, enabling an independent  control of the three axes, not
reached within the first order expansion.\\
\indent From  the  $\Lambda_1$ eigenvalues  we  derive  that the  rectified
 magnetic field produces an energy splitting described by an effective
 $\Omega_L$ Larmor precession frequency 
\begin{equation}
\Omega_{L}=\sqrt{\omega_{0x}^2+\widetilde{\omega}_{0y}^2+ \widetilde{\omega}_{0z}^2},
\label{eq:LarmorFreq}
\end{equation}
where for $p$ even
\begin{eqnarray}
\widetilde{\omega}_{0y}&=&J_0(\xi)\omega_{0y}+ J_p(\xi)\Omega_t\cos(\Phi), \nonumber \\
\widetilde{\omega}_{0z}&=&J_0(\xi)\omega_{0z},
\end{eqnarray}
and for $p$ odd
\begin{eqnarray}
\widetilde{\omega}_{0y}&=&J_0(\xi)\omega_{0y}, \nonumber \\
\widetilde{\omega}_{0z}&=&J_0(\xi)\omega_{0z}+ J_p(\xi)\Omega_t\sin(\Phi).
\end{eqnarray}
\indent    Eqs.~\eqref{eq:effectivefield}    and~\eqref{eq:LarmorFreq}
evidence  the  triaxial spatial  response  to  the external  drivings,
equivalent to an anisotropic non-linear gyromagnetic ratio.  

Generalizing the  analysis of~\cite{GolubLamoreaux_94}, the
temporal   evolution   of   the   atomic   coherences~\cite{BevilacquaSuppl20} 
for  an  initial  state  prepared  in  a
$\sigma_x$ eigenstate is
\begin{equation}
  \label{eq:sx:t:fin}
  \langle \sigma_x(t) \rangle  = 
  (1 - \frac{h_x^2}{\Omega_{L}^2}) \cos ( \Omega_{L} t ) + \frac{h_x^2}{\Omega_{L}^2}.
\end{equation}
 and contains only a precession at the $\Omega_L$ frequency.  Instead
$\langle \sigma_{y,z}(t) \rangle$
contain oscillations  also at
harmonics of the $\omega$ frequency.  


 \begin{figure}[t!!]
   \centering
   \includegraphics [angle=0, width= \columnwidth] {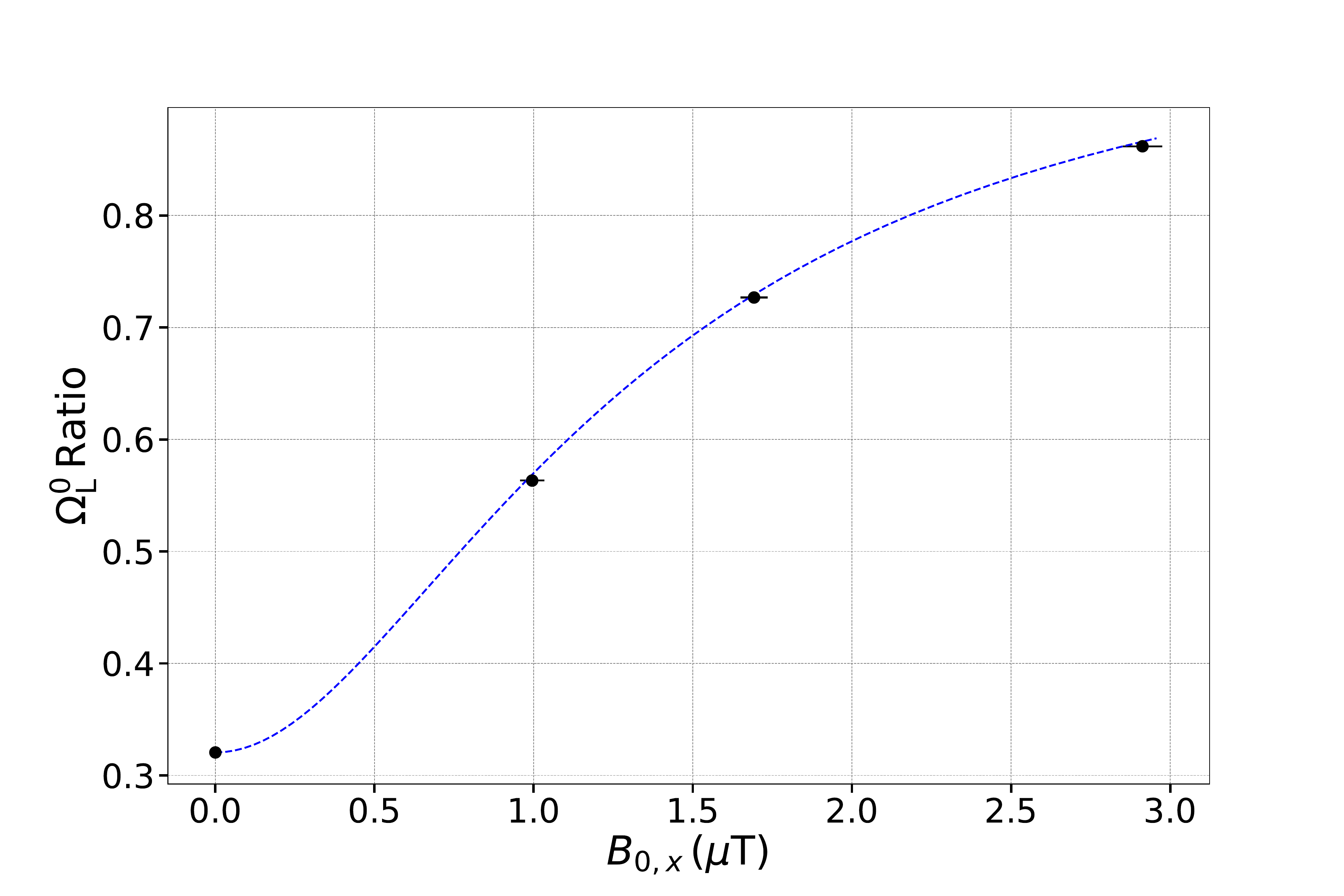}
    \caption{(Color online) $B_{0x}$ calibration data obtained for the
      dressed   $B_t=0$   case      at
      $\omega_{0z}/2\pi=5.979(2)$  kHz and $\omega/2\pi=30$  kHz. The ratio of
      the $\Omega_{L}^0$ measurements  at the $\xi=0$
      and  $\xi=1.833(5)$ values vs $B_{0x}$  is  marked  by the
      black  dots.  The fit  to  the  theoretical
      predictions determines  the $B_{0x}$  scale at the  four percent
      precision  level  given  by   the  horizontal  error  bars.  The
      $B_{0x}=0$ data point provides the $\xi$ calibration. } 
  \label{fig.calibration}
\end{figure} 

The  quantum
control   flexibility  associated  to   the  tuned-dressing is tested using the optical
magnetometric  apparatus of Ref.~\cite{biancalana_apb_16}.  The vapour
caesium sample is  pumped to the $F_g=4$ ground
hyperfine state  by the D1 line  and optically probed on  the D2 line.
The pump laser  propagates along the $x$ direction  of the oscillating
dressing field. The probe  laser along that direction monitors
the   atomic   evolution   given  by   Eq.~\eqref{eq:sx:t:fin}.    The
polarization of the transmitted probe  laser is analyzed by a balanced
polarimeter.   We  operate   in  a  Bell-Bloom-like  configuration  by
applying to the D1 pumping laser a wide-range periodic modulation with
frequency $\omega_M$.  This modulation creates also  the repumper from
the $F_g=3$ Cs ground state. By scanning $\omega_M$ around $\Omega_L$,
the  polarimetric signal  is analyzed  in order  to derive  the atomic
magnetic resonance  with a 20  Hz HWHM linewidth due  to spin-exchange
relaxation  and  probe  perturbations.  This system reaches   an    accuracy    at   the    Hz
level~\cite{bevilacqua_pra_12} for  frequency measurements. 

 A static magnetic  field is applied in a direction  of the $xz$ plane
 at  a variable  angle from  the $z$  axis.  Essential  components are
 three large  size, mutually orthogonal Helmholtz pairs,  here used to
 to lock the $B_{0x}$ and $B_{0z}$ field components to desired values,
 in the range $1-4$ $\mu$T ($\omega_{0x}/2\pi,\omega_{0z}/2\pi$ in the
 range  3-15  kHz),  and  to  compensate  the  $y$  component  of  the
 environmental  magnetic  field. Five quadrupoles  coils
 compensate the field gradients at the nT/cm level.

 We  operate  with  dressing  frequency  $\omega/2\pi=9-30$  kHz,  and
 $p=1-3$  values.  The  two  oscillating  rf fields  are  produced  by
 different  coils  driven by  phase-locked  waveform generators.   The
 $B_d$  field  is  generated by  a  long solenoidal  coil
 external  to the  magnetometer core.  The  $B_t$ field  is
 produced by  a separate  Helmholtz coil pair.  The  $B_d$  and $B_t$
 values may be  derived from geometry and current of  the coils at the
 few percent level.

\begin{figure}[]
   \centering
    \includegraphics [angle=0, width= \columnwidth] {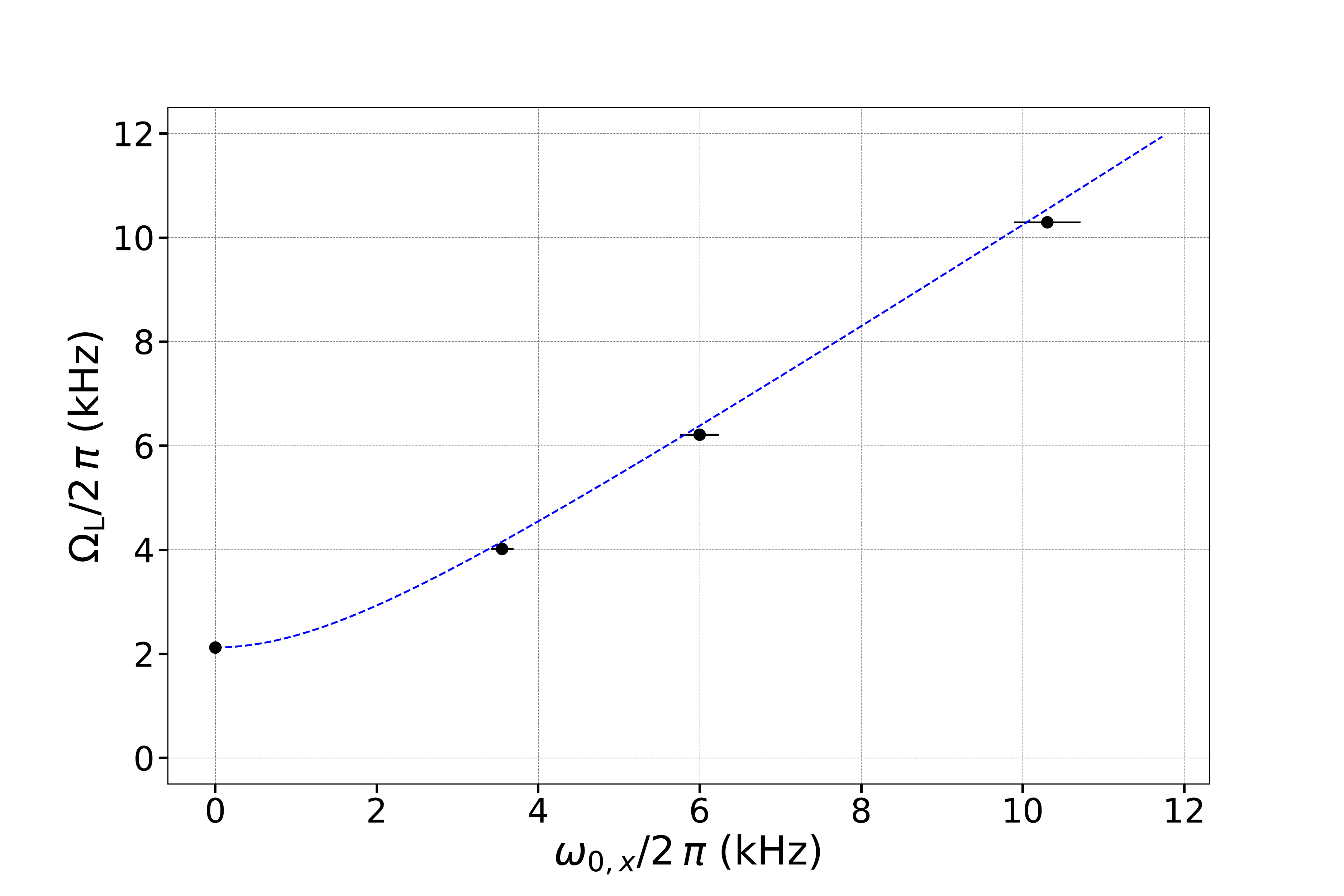}
    \caption{(Color online)  $\Omega_{L}$ vs $\omega_{0x}$ as
      a  test of  the spin  anisotropy,  for an  applied $p=1$  tuning
      field, $\Omega_t/2\pi=0.354(3)$ kHz,  and other parameters as in
      Fig.~\ref{fig.calibration}.  Blue  dots for data  and continuous
      line  for  the  theoretical  predictions.  Error  bars  two  per
      thousand on the frequency, and four percent on the $\omega_{0x}$
      scale.  } 
  \label{fig.anisotropy}
\end{figure}

For a higher precision determination  of $B_d$ and $B_{0x}$ we use the
following precession law in the $B_t=0$ case: 
\begin{equation}
\Omega^0_{L}= \sqrt{\omega_{0x}^2+\omega_{0z}^2J_0(\xi)^2}.
\label{eq:bareprecession}
\end{equation}   
For $B_{0x}=0$, a fit of the  $\Omega^0_{L}$ vs $\xi$ data determines 
  the dressing parameter at the three per thousand precision level. In
  order to  determine $B_{0x}$, we  measure $\Omega^0_{L}$ vs
  this  transverse   static  field  for  the  $\xi$   values  $0$  and
  $\approx1.83$ maximising  the $J_0$ slope.  A fit of their  ratio to
  the above precession  predictions, as in Fig.~\ref{fig.calibration},
  allows us to derive the $B_{0x}$ value at the four percent precision
  level.  In addition  the fit  determines that  the  applied $B_{0x}$
  field contains a three percent component along the $z$ axis.\\ 
\begin{figure}[t!!]
   \centering
    \includegraphics [angle=0, width= \columnwidth] {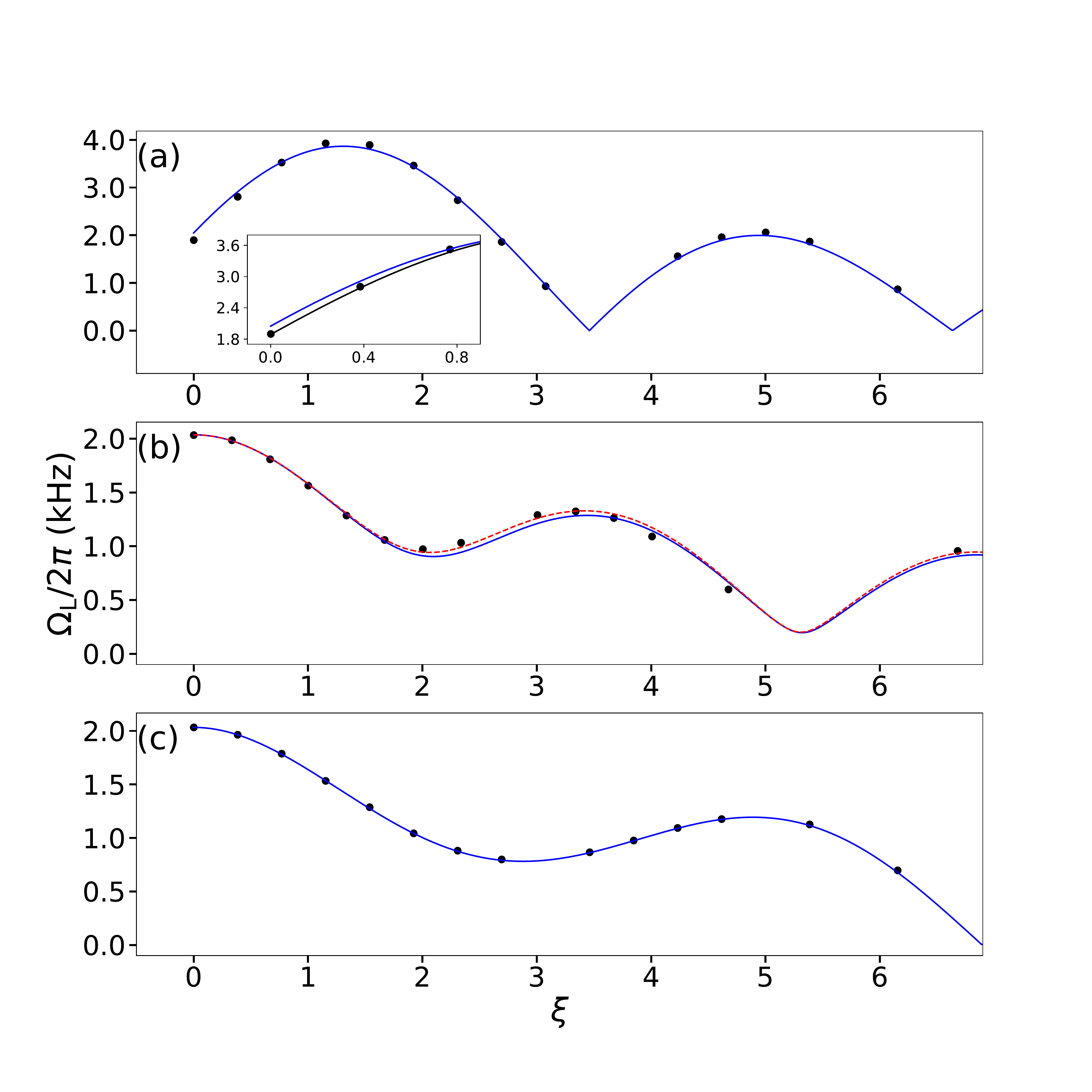}
    \caption{(Color online) Absolute $\Omega_{L}$ frequency vs 
   $\xi=\Omega_d/\omega$ scanning $\Omega_d/2\pi$ in the $(0,200)$ kHz
   range,     for     $\omega_{0z}/2\pi     =2.040(1)$    kHz,     and
   $\omega_{0x}=\omega_{0y}=0$. Parameters  $p$, $\Phi$, $\omega/2\pi$
   and $\Omega_t/2\pi$,  both in kHz: in  (a) $[1, \pi/2,9,4.97(25)]$;
   in (b)  $[2,0,10,2.23(10)]$; in (c)  $[3,\pi/2,9,4.25(20)]$.  Black
   dots for data. Error bars on the frequency at two per thousand, and
   on $\xi$ at the three per thousand level, both smaller than the dot
   size.  Theory in continuous blue lines from perturbative treatment; in (a)
   black line from the numerical analysis, in (b) dashed red one for a
   refined $\Omega_t/2\pi$ value, see text.  } 
  \label{fig.fields}
\end{figure} 
\indent In    order     to    verify  the $\Omega_{L}$ dependence on
the  quantum handles, we operate   with   the   $\omega_{0x}=\omega_{0y}=0$, where 
the   precession   frequency   becomes
$\Omega_{L}=\omega_{0z}J_0(\xi)+\Omega_tJ_p(\xi)\sin(\Phi)$
for                   $p$                   odd,                   and
$\Omega_{L}=\sqrt{\left[\Omega_tJ_p(\xi)\cos(\Phi)\right]^2+\left[\omega_{0z}J_0(\xi)\right]^2}$
for $p$  even.  The three  panels of Fig.~\ref{fig.fields}  report the
measured   (black    dots)   and   theoretical    (continuous   lines)
$\Omega_{L}$ absolute  values vs the  $\xi$ dressing
for  different   combinations  of  the  $p$,   $\Phi$  and  $\Omega_t$
parameters. Their  values are chosen  in order to maximise  the atomic
response tuning.   Panels (a) and  (c) deal with the odd $p=1,3$  values
where the $J_1$  and $J_3$ Bessel functions play the  key role for the
$\xi$  dependence.  Panel  (b)  dealing  with  the  even $p=2$   case
evidences  the  $J_2$  function  role  for  $\Omega_{L}$.  An
important  result  of the  $p=1$  plot  (a)    is  the possibility  of
increasing  the Larmor  frequency,  a feature  not  accessible to  the
single irradiation  configuration. The  odd harmonic cases allow for a sign  
change for the Larmor frequency, showing up
as a slope change in the plot (a) measured absolute value. A sign change occurs
also  in the  single  dressing  case, with  its  $J_0$ dependence  and
cylindrical symmetry.  Notice  that  in  Fig.~\ref{fig.fields}  the  perturbative
treatment   is  not  valid   for  the   $\xi\le\omega_{0z}/\omega,  \;
\Omega_t/\omega$ values, $\approx 0.5,\approx 0.4$ respectively.  The $\Omega_{L}$
value at $\xi=0$  is determined  by treating  $\Omega_t$ as  the dressing
field. A numerical analysis of the spin evolution, as in black line of
the panel (a) inset, leads to a better agreement with the data.\\ 
\begin{figure}[t!!]
   \centering
    \includegraphics [angle=0, width= \columnwidth] {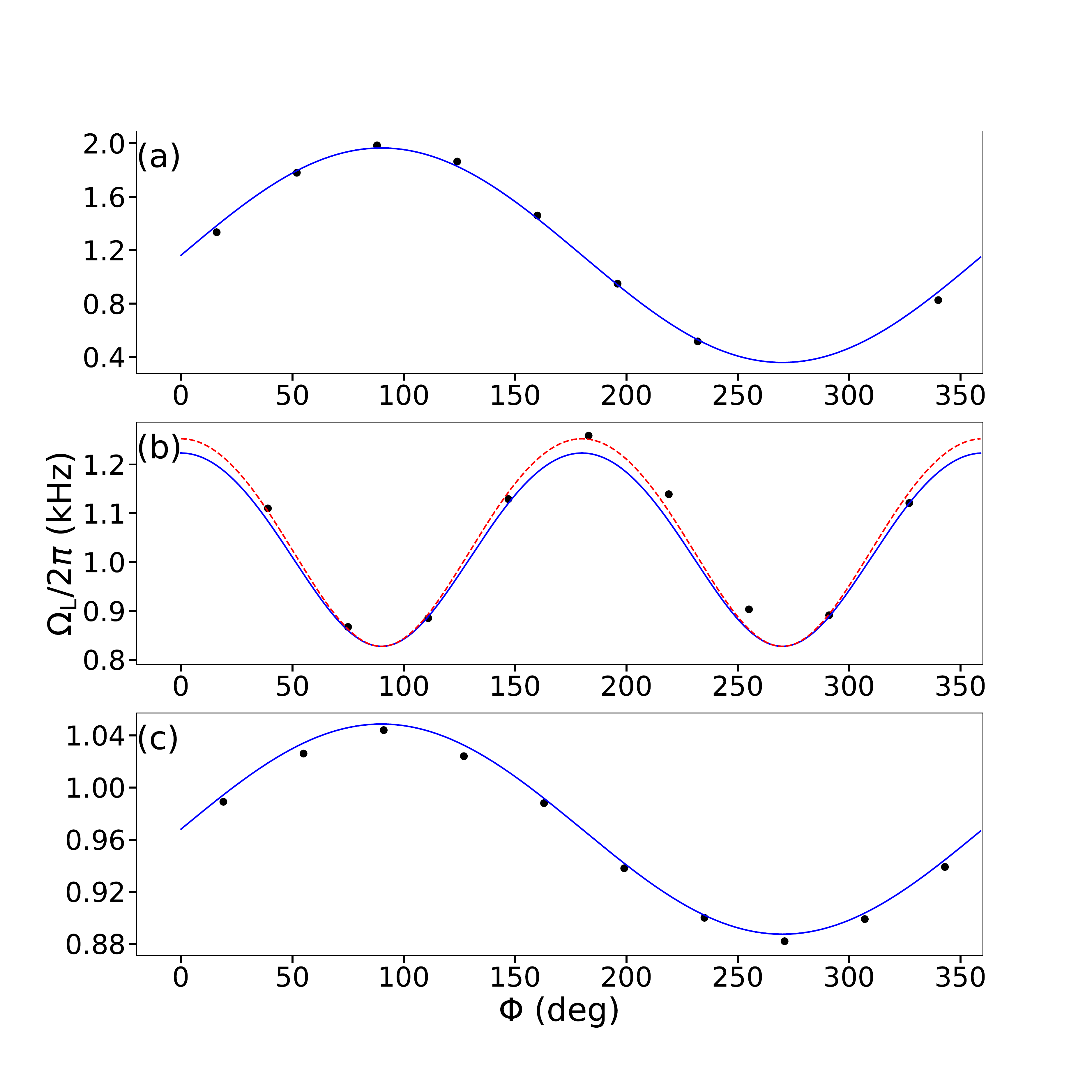}
    \caption{(Color online) $\Omega_{L}$  vs the $\Phi$ phase
      difference,       for       $\omega_{0x}=\omega_{0y}=0$      and
      $\omega_{0z}/2\pi=2.040(1)$   kHz.    Parameters   $p$,   $\xi$,
      $\omega/2\pi$   and  $\Omega_t/2\pi$,  both   in  kHz:   in  (a)
      $[1,1.38,20,1.49(8)]$;  in  (b)  $[2,3.83,10,2.23(12)]$; in  (c)
      $[3,1.54,9,1.23(6)]$.  Black  dots  data.  Error  bars  two  per
      thousand  on   the  frequency,  and   of  one  degree   for  the
      phase. Theoretical predictions given  by the blue continuous; 
      in (b) also by the red  dashed line for a refined $\Omega_t/2\pi$ value,
      both  presented in the text. } 
  \label{fig.phases}
\end{figure}
\indent     Fig.     \ref{fig.phases}     reports     the   
$\Omega_{L}$ dependence  on the $\Phi$ phase  with theoretical  predictions  given by  the  continuous lines.  In
panels (a) and (c) for odd harmonics, the data follows a sine profile,
with amplitudes given by $J_1(\xi)$ and $J_3(\xi)$. In
panel   (b)  for  an   even  harmonic,   the  variation   follows  the
squared-cosine   profile  with   amplitude   set  by   
$J_2(\xi)$. These  results confirm the usefulness of  the $\Phi$ phase
as an additional tuning-dressed parameter.    The   theory-data agreement for 
Figs.~\ref{fig.fields}   and  \ref{fig.phases} relies  
on  $\Omega_t$ precise  determination of the  tuning field
amplitude. The theoretical  analysis shows that for the  $p$ odd cases
the fit quality remains  constant for $\Omega_t$ variations within the
error  bar. Instead  for the  $p=2$  even case  of panel  (b) in  both
figures,  a $\Omega_t$  scaling up  by four  percent produces  the red
dashed  lines   with  a   better  data-theory
agreement.\\
 \indent In  order to  test the  full triaxial  anisotropy  Larmor frequencies
exploiting our  $x$  axis pump/probe geometry, we
modify the spin spatial evolution applying different $B_{0x}$ magnetic
fields for a fixed $B_{0z}$ value.  With this  tilted static field  the  $\Omega_L$ measurements
probe the  triaxial spatial dependence.   Fig.~\ref{fig.anisotropy}   reports  those measurements  as a  function of
the $B_{0x}$ value.  A precise theoretical  analysis requires
the  determination of  the applied  $\Omega_t$ field, derived here from   the above $B_{0x}=0$
measurements  with static  field along  the $z$ axis.  The continuous line  of figure shows
the excellent comparison  with the $p=1$ theory, confirming the quantum system anisotropy.\\ 
\indent For  magnetic    resonance,    the    condition    of
$\Omega_L$  exceeding  the  unperturbed one,  not  obtainable
using a single  dressing field, shifts the spin  resonant frequency to
higher frequencies where the detection sensitivity increases. 
 For  the handle of field inhomogeneities,  the $J_0(\xi)=0$ dressing
  (or  the magic  dressing of  Ref.\cite{ZanonArimondo_12}) eliminates
  the static  interaction dependence.  In our  scheme, the detrimental
  effects        caused       by       the        $\xi$       dressing
  inhomogeneities~\cite{Swank_pra_18} are  greatly reduced by operating
  at    a     small    $d\Omega_{L}/d\xi$    and    arbitrary
  $\Omega_{L}$.     In    magnetometry    applications    the
  $\Omega_{L}$     frequency     was    made     deliberately
  position-dependent     by      means     of     an     inhomogeneous
  $\xi$~\cite{biancalana_prappl_19,Bevilacqua_apl_19}.  Remarkably, in
  our scheme a space dependent $\Omega_{L}$ may be introduced
  by means of  a $B_t$ inhomogeneity, easier to  implement and control
  since  $B_t  \ll  B_d$.   Finally  the  tuning  field  $\Phi$  phase
  dependence could  be applied to complement  the amplitude dependence
  in the magnetometric detection of weakly conductive material
  targets~\cite{marmugi_apl_19,Deans:2020isr}. \\
\indent 
The basic tuning-dressed  mechanism is the interference in the
excitation  produced by  the two  harmonic rf  fields and  enhanced by
their   low-harmonic   order.     Such   interference   was   examined
in~\cite{TsukadaTomishima81} within a Green function approach. 
The
harmonic mixing of the biharmonic driving originates a rectified field
that  modifies   the  system  eigenenergies   and  eigenstates.   This
nonlinear  rectification process  borrows strength  from  the dressing
field  and is  associated  with  high order  light-shifts  due to  the
biharmonic     driving.      The     rectification    and     harmonic
mixing~\cite{GoychukHaenngi_98}     and    the     split    biharmonic
driving~\cite{LebedevRenzoni_09},   widely  investigated   within  the
quantum  ratchet  topic~\cite{HaenggiMarchesoni_09}, present  features
similar to  our investigation.  Those  systems deal with  the external
degrees of freedom, while our work examines the internal ones. However
the symmetries widely applied in quantum ratchets could represent a
tool for exploring the generation of rectified magnetic fields. \\
\indent The spin individual spatial components and their signs are not
accessible  to our experimental  investigation. A  direct test  of the
spatial anisotropy  can be obtained in a  critical dressing experiment
as in~\cite{HarocheCohenTannoudji_prl_70_2} with  spin exchange of the
transverse  magnetization  along  the  $x,y$ axes.  Playing  with  the
different  tuning response for  the two  investigated spins,  the spin
collapse in one direction and the enhancement in a different direction
will find their perfect testbed and also new
applications. \\
\indent  The   authors  thank  D.    Ciampini  and  F.    Renzoni  for
constructive    comments   and    criticism    on   the    manuscript.
\bibliographystyle{ieeetr}
\bibliography{biblioDressing_TZW}

\widetext
\clearpage

\begin{center}
\textbf{\large Supplemental Material: Harmonic 
  tuning of an atomic dressed system in magnetic resonance}
\end{center}
\setcounter{equation}{0}
\setcounter{figure}{0}
\setcounter{table}{0}
\setcounter{page}{1}
\makeatletter
\renewcommand{\theequation}{S\arabic{equation}}
\renewcommand{\thefigure}{S\arabic{figure}}
\renewcommand{\bibnumfmt}[1]{[S#1]}

\section{Spin 1/2 system}
\subsection{$U_I$ operator derivation}

The  manipulations  of the  Pauli  matrices  are  performed using  the
identities 
\begin{align}
\label{prop:Pauli:matr:1}
\e^{i \, \theta \, \mathbf{u} \cdot \boldsymbol{\sigma}/2}
   & =
\cos(\theta/2)   \mathbbmss{1}  +  i   \sin(\theta/2)\mathbf{u}  \cdot
\boldsymbol{\sigma} \\
\label{prop:Pauli:matr:2} 
\e^{i \, \theta \, \mathbf{u} \cdot \boldsymbol{\sigma}/2}
\mathbf{a}\cdot \boldsymbol{\sigma}\e^{-i \, \theta \, \mathbf{u} \cdot \boldsymbol{\sigma}/2}
& =
\cos\theta \,\mathbf{a}\cdot \boldsymbol{\sigma} - 
\sin\theta\,   (    \mathbf{u}     \times    \mathbf{a}    )    \cdot
\boldsymbol{\sigma} \\ \nonumber
& \phantom{=}+ (1-\cos\theta) ( \mathbf{u} \cdot \mathbf{a} ) ( \mathbf{u}\cdot 
\boldsymbol{\sigma} ) 
\end{align}
where $\boldsymbol{\sigma}$  denotes the vector of  the Pauli matrices
($  \boldsymbol{\sigma} =  ( \sigma_x,  \sigma_y, \sigma_z  )$), while
$\mathbf{u}$ and  $\mathbf{a}$ are two  vectors satisfying $\mathbf{u}
\cdot  \mathbf{u}  = 1  $,  and  $\mathbbmss{1}$  is the  $2\times  2$
identity  matrix.  It is  a  textbook  exercise  to demonstrate  these
formulas starting from the property of the Pauli matrices 
\[
(\mathbf{a} \cdot \boldsymbol{\sigma}) 
(\mathbf{b} \cdot \boldsymbol{\sigma}) = 
\mathbf{a} \cdot \mathbf{b} 
+ i (\mathbf{a} \times \mathbf{b} ) \cdot \boldsymbol{\sigma}.  
\]

Given the atomic Hamiltonian of Eq. (1) of the main text, the  $U_0$
dressing evolution is expanded as  by
\begin{equation} 
U_0=e^{-i\varphi(\tau)\sigma_x/2}=\cos(\varphi(\tau)/2) \mathbbmss{1}-\sin(\varphi(\tau)/2)\sigma_x. 
\end{equation}
The $U_I$
interaction evolution  is given by
\begin{equation}
  \label{eq:UI}
  \begin{split}
  i\dot{U}_I & = 
  U_0^{\dagger}
  \left[\frac{1}{2
      \omega} \boldsymbol{\omega}_0 \cdot \boldsymbol{\sigma}+
    \frac{\Omega_t}{2\omega}\cos(p\tau+\Phi)\sigma_y\right]
  U_0 
\; \; U_I \\
& =  \frac{1}{2\omega}[ \cos(\varphi) 
\boldsymbol{\omega}_0 \cdot \boldsymbol{\sigma}
-\sin(\varphi)(\mathbf{\hat{x}}\times\boldsymbol{\omega}_0) \cdot 
  \boldsymbol{\sigma} \\ 
  &\phantom{====}  +(1-\cos(\varphi))\omega_{0x}\sigma_x ]\; U_I \\
  & \phantom{=}+ \frac{\Omega_{t}}{2\omega}\cos(p\tau+\Phi)\left[\cos(\varphi)\sigma_y
    -\sin(\varphi) \sigma_z \right]\;  U_I .
\end{split}
\end{equation} 
After some straightforward algebra one obtains the expression reported
in the main text 
where the explicit form of the $g_y,g_z$ functions is
\begin{eqnarray*}
  g_y &=& + \frac{\omega_{0y}}{2\omega}\cos(\varphi)+
  \frac{\omega_{0z}}{2\omega}\sin(\varphi)
  +\frac{\Omega_t}{2\omega}\cos(\varphi)\cos(p\tau+\Phi), \\
  g_z &=& -\frac{\omega_{0y}}{2\omega}\cos(\varphi)+
  \frac{\omega_{0z}}{2\omega}\cos(\varphi) - 
  \frac{\Omega_t}{2\omega}\sin(\varphi)\cos(p\tau+\Phi).
\end{eqnarray*}

The first order operators needed in the Floquet-Magnus expansion are
given explicitly as
\begin{equation}
  \label{eq:F1:L1}
  \begin{split}
    \Lambda_1 &= \frac{1}{2\pi} \int_0^{2\pi} A(\tau) \mathrm{d}\,\tau \\
    P_1(\tau) &=  \int_0^{\tau} A(\tau') \mathrm{d}\,\tau' -\tau \Lambda_1.
  \end{split}
\end{equation}

The involved time integrals are reported in the Appendix.

\subsection{Spin coherences}
\indent For the lowest order determination of the atomic coherences, we approximate $\e^{-iP_1(\tau)}$
 by the identity matrix (see the Appendix).  The time  evolution operator becomes
$U(\tau)  \approx  \e^{-  i \varphi(  \tau)  \;  \sigma_x/2} \,  \e^{-i   \Lambda_1 \tau}
 $. Therefore introducing  the
dimensional time, we obtain for the $\sigma_x$ operator    
\begin{equation}
  \label{eq:sx:t:ini}
  \begin{split}
    \sigma_x(t)&= U(t)^{\dagger} \sigma_x U(t) \\ 
    &= \e^{i \Lambda_1\omega t } \e^{i \varphi(t) \sigma_x/2} \sigma_x 
    \e^{-i \varphi(t) \sigma_x/2} \e^{-i \Lambda_1\omega t } \\
    &= \e^{i \Lambda_1\omega t } \sigma_x \e^{-i \Lambda_1\omega t }\\
    &= \e^{i \mathbf{h}\cdot\boldsymbol{\sigma} t/2 } \sigma_x 
    \e^{-i \mathbf{h}\cdot\boldsymbol{\sigma} t/2 } \\
    &=     \e^{i    \Omega_L     t    \,     (\mathbf{h}/    \Omega_L)
      \cdot\boldsymbol{\sigma}/ 2 } \sigma_x 
    \e^{-i         \Omega_L          t         \,         (\mathbf{h}/
      \Omega_L)\cdot\boldsymbol{\sigma} /2 } 
\end{split}
\end{equation}
where the vector $\mathbf{h}$ is reported  in the main text and from the
last   line   one   can   see   that   it   is   possible   to   apply
\eqref{prop:Pauli:matr:2} with $\theta=\Omega_L t$ and $\mathbf{u} = \mathbf{h}/\Omega_L$
being $\Omega_L$ the modulus of $\mathbf{h}$. The final result is
\begin{equation}
\begin{split}
  \label{eq:sx:t:fin}
  &\sigma_x(t)=  \cos ( \Omega_{L} t ) \sigma_x - \sin(\Omega_{L} t ) 
    \left( \frac{h_z}{ \Omega_L} \sigma_z - \frac{h_y}{ \Omega_L}\sigma_y\right) + \\
    &\phantom{= } \left(1-\cos (\Omega_{L} t )\right) \frac{h_x}{ \Omega_L} 
    \left( \frac{h_x}{ \Omega_L} \sigma_x + \frac{h_y}{ \Omega_L} \sigma_y + \frac{h_z}{ \Omega_L} \sigma_z  \right).
  \end{split}
\end{equation}
If the initial state  is prepared in a $\sigma_x$ eigenstate, the $x$ axis coherence becomes that reported within the main text.\\
\indent Repeating the derivation for the $y$ axis we obtain
\begin{equation}
  \label{eq:sy:t:fin}
  \begin{split}
    \langle \sigma_y(t) \rangle &= 
    \left[ 
      \frac{h_y}{ \Omega_L} \sin(\varphi(t))+ 
      \frac{h_z}{ \Omega_L} \cos(\varphi(t)) 
    \right]
    \sin(\Omega_{L} t ) +\\
    &
    \left[
      \frac{h_x h_y}{ \Omega_L^2} \cos( \varphi(t)) - 
      \frac{h_x h_z}{ \Omega_L^2} \sin(\varphi(t)) 
    \right]
    (1- \cos ( \Omega_{L} t )).
 \end{split}
\end{equation}
For the $z$ axis we obtain  
\begin{equation}
  \label{eq:sz:t:fin}
  \begin{split}
   & \langle \sigma_z(t) \rangle = 
    \left[    \frac{h_z}{ \Omega_L}    \sin(\varphi(t))    -    \frac{h_y}{ \Omega_L}    \cos(\varphi(t))
    \right]\sin(\Omega_{L} t ) +\\
    &\phantom{=}
    \left[ \frac{h_x h_z}{ \Omega_L^2} \cos( \varphi(t)) + \frac{h_x h_y}{ \Omega_L^2} \sin(\varphi(t)) \right]
    (1- \cos ( \Omega_{L} t )). \\
  \end{split}
\end{equation}

\subsection{Tuning field in an arbitrary direction}
We derive the  spin effective field for the case of  a tuning rf field
oriented in an arbitrary  direction and having $(X_t,Y_t,Z_t)$ spatial
components  with $(q,p,r)$ harmonic  temporal dependencies,  the index
$p$ being  assigned to the  $y$ axis as  in the main text,  and phases
$(\Phi_x,\Phi_y,\Phi_z)$,  respectively. The  spin 1/2  interaction is
described by the following Hamiltonian:
\begin{eqnarray}
H= &\frac{Y_t}{2\omega}\cos(p\tau+\Phi_y)\sigma_y\nonumber \\
+&\frac{X_t}{2\omega}\cos(q\tau+\Phi_x)\sigma_x++\frac{Z_t}{2\omega}\cos(r\tau+\Phi_z)\sigma_z.
\label{HamiltArb:eq}
\end{eqnarray}
\indent  We repeat  the Floquet-Magnus  expansion for  the interaction
operator.  The analysis  for  the $\mathbf{h}$  first order  effective
magnetic field  introduced by Eq.  (8) of the  main text leads  to the
following components  associated to  different even/odd values  of the
harmonic coefficients
\begin{eqnarray}
 &h_x&=    \omega_{0x}, \\
 & h_y&= J_0(\xi)\omega_{0y}\nonumber \\
        &+& J_p(\xi)Y_t\cos(\Phi_y)  \;(p, r) \;\textup{even},\nonumber \\
        &-& J_r(\xi)Z_t\sin(\Phi_z) \;(p, r)  \;\textup{odd}, \nonumber \\
        &+&J_p(\xi)Y_t\cos(\Phi_y)- J_r(\xi)Z_t\sin(\Phi_z), \; p \;\textup{even}, \;\textup{and} \;  r  \;\textup{odd}, \nonumber \\
        &+&0 \; p \; \textup{odd}, \; \textup{and} \;r  \;\textup{even}   ,
\end{eqnarray}
\begin{eqnarray}
  \label{eq:gLandrez}
  & h_z&= J_0(\xi)\omega_{0z}\nonumber \\
        &+& J_r(\xi)Z_t\cos(\Phi_z)  \;(p, r) \;\textup{even}, \nonumber \\
        &+& J_p(\xi)Y_t\cos(\Phi_y)  \;(p, r) \;\textup{odd}, \nonumber \\
        &+&0  \; p \;\textup{even} \;\textup{and} \;r  \;\textup{odd},   \nonumber \\
        &+& J_p(\xi)Y_t\sin(\Phi_y)+ J_r(\xi)Z_t\cos(\Phi_z) \; p \;\textup{odd} \;\textup{and} \;  r  \;\textup{even}.  
\end{eqnarray}

\section{Systems with a higher spin}
While the main text examines the  simple case of a two level atom, the
present analysis confirms the validity also for a higher spin
system, as  for the caesium  ground state experimentally  tested.   \\
\indent The evolution of  atomic magnetization $\mathbf{M}$, i.e., the
mean value of a quantum operator, in an external field $\mathbf{B}$ is
described by the Bloch equations
\begin{equation}
\mathbf{\dot{M}} = \gamma \mathbf{B}
\times \mathbf{M}.
\end{equation}  
By examining the $B_{0x}= B_{0y}=0$ case the equation  for the atomic magnetization
 in presence of a magnetic field is
\begin{equation}
  \label{eq:Larmor:adim}
  \frac{d\, \mathbf{M}}{d\, \tau} = 
  \left[ \frac{\Omega_d}{\omega} \cos (\tau)\, L_x +
    \frac{\Omega_t}{\omega} \cos( p \tau + \Phi) \, L_y+
    \frac{\omega_{0z}}{\omega} L_z 
  \right] \mathbf{M}  
\end{equation}
where $L_j$ with $(j=x,y,z)$ are the 
$L=1$  angular  momentum operator matrices.\\
\indent Using the  perturbation theory  we factorize the
time  evolution operator  $U(\tau)$, i.e.,  $\mathbf{M}(\tau)  \equiv U(\tau)
\mathbf{M}(0)$, in the interaction representation as
\begin{eqnarray}
  \label{eq:U:repr:int}
  U(\tau) &=& e^{\left[\xi\ \sin \tau \, L_x\right]}\, U_I(\tau)  \nonumber \\
  &=&
  \begin{pmatrix}
    1 &0 &0 \\
    0 &\cos \varphi(\tau) & - \sin \varphi(\tau) \\
    0 & \sin \varphi(\tau) & \cos \varphi(\tau)
  \end{pmatrix} U_I(\tau),
\end{eqnarray}
and we  obtain  the following
dynamical equation for $U_I(\tau)$:
\begin{equation}
  \label{eq:UI:3D}
  \begin{split}
  \frac{d\, U_I}{d\, \tau} & = \left[ 
    \left(      \frac{\omega_{0z}}{\omega}       \sin      (\varphi)     +
      \frac{\Omega_t}{\omega}  \cos(\varphi)  \,  \cos(p\tau  +  \Phi)
    \right) L_y + \right. \\
    & \phantom{= \left[\right.}\left. \left( \frac{\omega_{0z}}{\omega} \cos (\varphi)-
        \frac{\Omega_t}{\omega}  \sin  (\varphi)  \,  \cos(p\tau  +  \Phi)\right) L_z 
    \right] U_I \\
    & \equiv \epsilon A(\tau) U_I.
\end{split}
\end{equation}
 \indent Because the matrix $A(\tau)$ is periodic $A(\tau  + 2\pi) = A(\tau)$, we follow the main text steps to write $U_I(\tau)$ as in Eq.~(4), and we introduce the    Floquet-Magnus expansion to  calculate the lowest order
terms.  For the first order perturbation we make use of  Eqs.~\eqref{eq:cos:sin:int} of the Appendix to obtain 
\begin{equation}
  \label{eq:F1}
  \Lambda_1 =
  \begin{cases}
    \frac{\omega_{0z}}{\omega} J_0(\xi)  \, L_z +  \frac{\Omega_t}{\omega} J_p(\xi)
    \cos \Phi\, L_y & p \; \mathrm{even}, \\
    \left[ \frac{\omega_{0z}}{\omega} J_0(\xi) + \frac{\Omega_t}{\omega} J_p(\xi)
      \sin \Phi \right] L_z & p \; \mathrm{odd}, 
  \end{cases}
\end{equation}
and 
\begin{eqnarray}
  \label{eq:Lambda1}
    P_1 &=& \left( \frac{\omega_{0z}}{\omega} f_2(\tau)  
      + \frac{\Omega_t}{\omega}f_3(\tau)\right) L_y \nonumber \\ 
      &+&
          \left( \frac{\Omega_t}{\omega} f_1(\tau) - 
        \frac{\Omega_t}{\omega}f_4(\tau)\right) L_z. 
\end{eqnarray}
The functions $f_i(\tau)$ are reported in the Appendix.\\
\indent  From   Eq.~\eqref{eq:F1}  we  
calculate the eigenvalues $(-\Omega_L,0,\Omega_L)$ with Larmor frequency
\begin{equation}
  \label{eq:fin:W_L}
  \Omega_{L} = 
  \begin{cases}
    \sqrt{\omega_{0z}^2 J_0^2(\xi)  + \Omega_t^2 J_p^2(\xi)\  \cos^2\Phi} & p\;
    \mathrm{even},\\
    \left|\omega_{0z} J_0(\xi) + \Omega_t J_p(\xi)\ \sin \Phi\right| & p\; \mathrm{odd}.
  \end{cases}
\end{equation}
These equations are  equivalent to those derived within  the main text
for a  two-level system. The present  approach can be  extended to the
Zeeman structure for higher spin systems.

\section{Appendix}

Exploiting the $J_n$ Bessel expansion
\begin{equation}
  \label{eq:bessel}
  \e^{i  \,  z  \sin  \theta}  =  \sum_{n=-\infty}^{+\infty}  J_n(z)\,
  \e^{i\, n \, \theta}
\end{equation}
the time integrals for the $\Lambda_1$ and $P_1$ derivation become
\begin{eqnarray}
  \label{eq:cos:sin:int}
    \int_0^{\tau} \cos( \varphi(\tau)) d \tau & = J_0(\xi) \tau + f_1(\tau), \nonumber  \\
   \int_0^{\tau} \sin( \varphi(\tau)) d \tau  & = f_2(\tau), \nonumber \\
    \int_0^{\tau} \cos( \varphi(\tau')) \cos( p \tau' +\Phi) d \tau & = \frac{1+(-1)^p}{2}\tau J_p(\xi) \cos\Phi  \nonumber \\
    & \phantom{=} + f_3(\tau), \nonumber \\ 
    \int_0^{\tau} \sin( \varphi(\tau')) \cos( p \tau' +\Phi) d \tau
    &= \frac{-1+(-1)^p}{2}\tau J_p(\xi) \sin\Phi  \nonumber \\
    & \phantom{=} + f_4(\tau).
\end{eqnarray}

These auxiliary $f_i$ functions are defined as 
\begin{align}
  \label{eq:f_i:def}
  f_1(\tau)  &=  \sum_{n=1}^{\infty} \frac{J_{2  n}(\xi)}{n  }  \sin  ( 2  n
    \tau), \nonumber \\
    f_2(\tau) &=  4 \sum_{n=0}^{\infty}  \frac{J_{2 n +  1}(\xi)}{ 2  n +1}
    \sin^2 ( (n+1/2)\tau),\nonumber \\
    f_3(\tau) &= \Re ( g(\tau) ),\nonumber \\
    f_4(\tau) &= \Im ( g(\tau) ),
\end{align}
where 
\begin{equation}
  \label{eq:g:def}
  \begin{split}
    g(\tau) = & e^{i  \Phi} \sum_{n \neq -p}\frac{J_n(\xi)}{i(n+p)} \left(
      e^{i(n+p)\tau} -1\right) +\\
      & e^{- i \Phi} \sum_{n \neq p}\frac{J_n(\xi)}{i(n - p)} \left(
        e^{i(n -p)\tau} -1\right). 
  \end{split}
\end{equation}
These functions, required in the evaluation 
 of $e^{-iP_1(\tau)}$, have a  limited and oscillating behaviour.  One can see  by inspection
that    $e^{-iP_1}    \approx    \mathbbmss{1}$   is    a    good
approximation.  

\end{document}